\documentclass[aps,prl,twocolumn,showpacs,preprintnumbers,superscriptaddress,floatfix]{revtex4}
\pdfoutput=1

\usepackage{graphicx}  

\usepackage{epsfig}

\usepackage{subfigure}

\usepackage{lineno}

\usepackage{rotating}

\usepackage{multirow}

\usepackage{dcolumn} 

\usepackage{afterpage}
\usepackage{graphics}
\usepackage{float}

\usepackage{verbatim}

\usepackage{color}

\usepackage{bm}

\usepackage{amssymb}

\usepackage[centertags]{amsmath}

\usepackage{indentfirst}

\usepackage{xspace}

\usepackage{enumerate}

\newcommand{\sk}    {Super-Kamiokande\xspace}

\newcommand{\ski}   {SK-I\xspace}
\newcommand{\skii}  {SK-II\xspace}
\newcommand{\skiii} {SK-III\xspace}

\newcommand{\ktyrs} {\ensuremath{\mathrm{kton\cdot yrs}}\xspace}
\newcommand{\ktyears} {\ensuremath{\mathrm{kton\cdot years}}\xspace}
\newcommand{\water} {\ensuremath{\mathrm{H_{2}O}}\xspace}
\newcommand{\mevc}  {\ensuremath{\mathrm{MeV}/c}\xspace}
\newcommand{\mevcc} {\ensuremath{\mathrm{MeV}/c^{2}}\xspace}

\newcommand{\p}     {\ensuremath{p}\xspace}
\newcommand{\n}     {\ensuremath{n}\xspace}

\newcommand{\pip}   {\ensuremath{\pi^+}\xspace}

\newcommand{\piz}   {\ensuremath{\pi^0}\xspace}

\newcommand{\anu}   {\ensuremath{\bar\nu}}

\newcommand{\nnupiz}{\ensuremath{\n\rightarrow\anu\piz}\xspace}
\newcommand{\pnupip}{\ensuremath{\p\rightarrow\anu\pip}\xspace}

\newcolumntype{.}{D{.}{.}{-1}}

\begin{document}




\title{A Search for Nucleon Decay via \nnupiz and \pnupip in Super-Kamiokande}

\newcommand{\AFFicrr}{\affiliation{Kamioka Observatory, Institute for Cosmic Ray Research, University of Tokyo, Kamioka, Gifu 506-1205, Japan}}
\newcommand{\AFFkashiwa}{\affiliation{Research Center for Cosmic Neutrinos, Institute for Cosmic Ray Research, University of Tokyo, Kashiwa, Chiba 277-8582, Japan}}
\newcommand{\AFFbu}{\affiliation{Department of Physics, Boston University, Boston, MA 02215, USA}}
\newcommand{\AFFbnl}{\affiliation{Physics Department, Brookhaven National Laboratory, Upton, NY 11973, USA}}
\newcommand{\AFFucd}{\affiliation{Department of Physics, University of California, Davis, Davis, CA 95616, USA}}
\newcommand{\AFFuci}{\affiliation{Department of Physics and Astronomy, University of California, Irvine, Irvine, CA 92697-4575, USA }}
\newcommand{\AFFcsu}{\affiliation{Department of Physics, California State University, Dominguez Hills, Carson, CA 90747, USA}}
\newcommand{\AFFcnm}{\affiliation{Department of Physics, Chonnam National University, Kwangju 500-757, Korea}}
\newcommand{\AFFduke}{\affiliation{Department of Physics, Duke University, Durham NC 27708, USA}}
\newcommand{\AFFfukuoka}{\affiliation{Junior College, Fukuoka Institute of Technology, Fukuoka, 811-0214, Japan}}
\newcommand{\AFFgmu}{\affiliation{Department of Physics, George Mason University, Fairfax, VA 22030, USA }}
\newcommand{\AFFgifu}{\affiliation{Information and Multimedia Center, Gifu University, Gifu, Gifu 501-1193, Japan}}
\newcommand{\AFFuh}{\affiliation{Department of Physics and Astronomy, University of Hawaii, Honolulu, HI 96822, USA}}
\newcommand{\AFFkanagawa}{\affiliation{Physics Division, Department of Engineering, Kanagawa University, Kanagawa, Yokohama 221-8686, Japan}}
\newcommand{\AFFkek}{\affiliation{High Energy Accelerator Research Organization (KEK), Tsukuba, Ibaraki 305-0801, Japan }}
\newcommand{\AFFkobe}{\affiliation{Department of Physics, Kobe University, Kobe, Hyogo 657-8501, Japan}}
\newcommand{\AFFkyoto}{\affiliation{Department of Physics, Kyoto University, Kyoto, Kyoto 606-8502, Japan}}
\newcommand{\AFFumd}{\affiliation{Department of Physics, University of Maryland, College Park, MD 20742, USA }}
\newcommand{\AFFmit}{\affiliation{Department of Physics, Massachusetts Institute of Technology, Cambridge, MA 02139, USA}}
\newcommand{\AFFmiyagi}{\affiliation{Department of Physics, Miyagi University of Education, Sendai, Miyagi 980-0845, Japan}}
\newcommand{\AFFnagoya}{\affiliation{Solar Terrestrial Environment Laboratory, Nagoya University, Nagoya, Aichi 464-8602, Japan}}
\newcommand{\AFFkmiopu}{\affiliation{Kobayashi-Maskawa Institute for the Origin of Particles and the Universe, Nagoya University, Nagoya, Aichi 464-8602, Japan}}
\newcommand{\AFFsuny}{\affiliation{Department of Physics and Astronomy, State University of New York, Stony Brook, NY 11794-3800, USA}}
\newcommand{\AFFniigata}{\affiliation{Department of Physics, Niigata University, Niigata, Niigata 950-2181, Japan }}
\newcommand{\AFFokayama}{\affiliation{Department of Physics, Okayama University, Okayama, Okayama 700-8530, Japan }}
\newcommand{\AFFosaka}{\affiliation{Department of Physics, Osaka University, Toyonaka, Osaka 560-0043, Japan}}
\newcommand{\AFFseoul}{\affiliation{Department of Physics, Seoul National University, Seoul 151-742, Korea}}
\newcommand{\AFFshizuokasc}{\affiliation{Department of Informatics in Social Welfare, Shizuoka University of Welfare, Yaizu, Shizuoka, 425-8611, Japan}}
\newcommand{\AFFskk}{\affiliation{Department of Physics, Sungkyunkwan University, Suwon 440-746, Korea}}
\newcommand{\AFFtohoku}{\affiliation{Research Center for Neutrino Science, Tohoku University, Sendai, Miyagi 980-8578, Japan}}
\newcommand{\AFFtokyo}{\affiliation{The University of Tokyo, Bunkyo, Tokyo 113-0033, Japan }}
\newcommand{\AFFipmu}{\affiliation{Kavli Institute for the Physics and Mathematics of the Universe (WPI), Todai Institutes for Advanced Study, University of Tokyo, Kashiwa, Chiba 277-8583, Japan}}
\newcommand{\AFFtokai}{\affiliation{Department of Physics, Tokai University, Hiratsuka, Kanagawa 259-1292, Japan}}
\newcommand{\AFFtit}{\affiliation{Department of Physics, Tokyo Institute for Technology, Meguro, Tokyo 152-8551, Japan }}
\newcommand{\AFFtsinghua}{\affiliation{Department of Engineering Physics, Tsinghua University, Beijing, 100084, China}}
\newcommand{\AFFwarsaw}{\affiliation{Institute of Experimental Physics, Warsaw University, 00-681 Warsaw, Poland }}
\newcommand{\AFFuw}{\affiliation{Department of Physics, University of Washington, Seattle, WA 98195-1560, USA}}
\newcommand{\AFFuam}{\affiliation{Department of Theoretical Physics, University Autonoma Madrid, 28049 Madrid, Spain }}

\AFFicrr
\AFFkashiwa
\AFFuam
\AFFbu
\AFFbnl
\AFFuci
\AFFcsu
\AFFcnm
\AFFduke
\AFFfukuoka
\AFFgifu
\AFFuh
\AFFkek
\AFFkobe
\AFFkyoto
\AFFmiyagi
\AFFnagoya
\AFFkmiopu
\AFFsuny
\AFFokayama
\AFFosaka
\AFFseoul
\AFFshizuokasc
\AFFskk
\AFFtokai
\AFFtokyo
\AFFipmu
\AFFtsinghua
\AFFwarsaw
\AFFuw
%

\author{K.~Abe}
\author{Y.~Hayato}
\AFFicrr
\AFFipmu
\author{T.~Iida}
\author{K.~Iyogi} 
\AFFicrr
\author{J.~Kameda}
\author{Y.~Koshio}
\AFFicrr
\AFFipmu
\author{Y.~Kozuma} 
\author{Ll.~Marti} 
\AFFicrr
\author{M.~Miura} 
\author{S.~Moriyama} 
\author{M.~Nakahata} 
\author{S.~Nakayama} 
\author{Y.~Obayashi} 
\author{H.~Sekiya} 
\author{M.~Shiozawa} 
\author{Y.~Suzuki} 
\author{A.~Takeda} 
\AFFicrr
\AFFipmu
% ...

\collaboration{The Super-Kamiokande Collaboration}
\noaffiliation

\date{\today}

\begin{abstract}
\vspace{0.1in} We present the results of searches for nucleon decay via
\nnupiz and \pnupip using data from a combined 172.8~\ktyears exposure of
Super-Kamiokande-I, -II, and -III.  We set lower limits on the partial
lifetime for each of these modes: $\tau_{\nnupiz}>1.1\times10^{33}$~years and
$\tau_{\pnupip}>3.9\times10^{32}$~years at 90\% confidence level.
\end{abstract}


\pacs{13.30.-a,11.30.Fs,12.60.Jv,14.20.Dh,29.40.Ka} 


\maketitle



Although there is strong theoretical support that nature can be described by
a grand unified theory (GUT)~\cite{Glashow:1974p1461,Salam:1974p1424}, there
is currently no direct experimental evidence.  One of the most powerful ways
to test grand unification is to look for proton (or bound neutron)
decay. Most GUTs have an unstable proton; in the absence of an observation,
setting experimental limits on the proton lifetime can provide useful
constraints on the nature of grand unified theories. Observation, on the
other hand, would be tantalizing evidence of new physics beyond the Standard
Model.

One of the more simple but interesting candidates for grand unification is
SO(10), where the Standard Model's SU(3), SU(2), and U(1) are contained
within the larger gauge group. The class of models based on SO(10)
unification generally make predictions for neutrino masses and mixing that
are broadly in accord with all known neutrino mixing
data~\cite{Goh:2004p279,Babu:2010p1621}. The minimal supersymmetric SO(10)
model with a {\bf{126}} Higgs field described in Ref.~\cite{Goh:2004p279} is
the particular motivation for the analysis presented here. In addition to
predicting neutrino mass and mixing in agreement with observations, it leaves
R-parity unbroken, which guarantees the existence of stable dark matter. For
some region of its allowed parameter space, this model predicts that the
dominant nucleon decay modes will be \pnupip and \nnupiz.


In this Letter, the fully-contained (FC) atmospheric neutrino data
({\it{i.e.}}, having activity only within the inner detector region and no
activity in the outer detector) collected during the first three running
periods of \sk (Super-K, SK) are analyzed in a search for both \pnupip
and \nnupiz: \ski (May 1996--Jul 2001, 1489.2 live days), \skii (Jan
2003--Oct 2005, 798.6 live days), and \skiii (Sep 2006--Aug 2008, 518.1 live
days). The combined dataset corresponds to an exposure of 172.8~\ktyrs.

The 50-kiloton (22.5~kton fiducial) Super-K water Cherenkov detector is
located beneath 1~km of rock overburden (2700 meters water equivalent) in the
Kamioka mine in Japan. Details of the detector design, calibration, and
simulations in \ski may be found in Ref.~\cite{Fukuda:2003p1333} and a
discussion of the reduced photo-sensor coverage in \skii may be found in
Ref.~\cite{Nishino:2011LPM}. In \skiii, the photo-sensor coverage is
recovered to the original 40\% level of \ski.

The efficiency of detecting nucleon decay occurring in the water is estimated
by Monte Carlo (MC) simulation. As discussed in detail in
Ref.~\cite{Nishino:2011LPM}, all nucleons in the \water molecule are assumed
to decay with equal probability, and Fermi motion, nuclear binding energy,
and meson-nuclear interactions in oxygen are taken into account.

The \nnupiz (\pnupip) decay mode results in a \piz (\pip) with mean momentum
460~\mevc (458.8~\mevc), smeared by the Fermi motion of nucleons bound in
oxygen, but uniquely determined for decays of the two free protons in the
water's hydrogen nuclei. The pion's momentum is also affected by nuclear
interactions as it travels through the nucleus; it may undergo scattering,
charge exchange, or absorption. Pion-nucleon interactions are carefully
simulated to reflect our understanding of the processes which may affect the
ability to detect pions in water.

Since the final state neutrino is undetected, these two modes of nucleon
decay are particularly challenging. It is not possible to develop a set of
selection cuts that will eliminate most of the background of atmospheric
neutrino interactions with a single pion and no other detected particles in
the final state. Instead, we select events that appear to have only a
single \piz or \pip, and perform a spectrum fit to their reconstructed
momentum distributions, respectively. In this method, the nearly
mono-energetic pions from nucleon decay would appear as a bump on top of the
broad distribution of pions from atmospheric neutrino
interactions. Atmospheric neutrino background events are simulated using the
NEUT neutrino interaction MC simulation~\cite{neut} with an
atmospheric neutrino flux calculated by Honda {\it et al.}~\cite{Honda:2007},
then passed through a GEANT-3-based~\cite{Geant3} custom detector simulation
that is described in detail in Ref.~\cite{Ashie:2005}.

\begin{figure*}[htb]
\subfigure[]{ 
   \includegraphics[width=0.4\textwidth,keepaspectratio=true,type=pdf,ext=.pdf,read=.pdf]{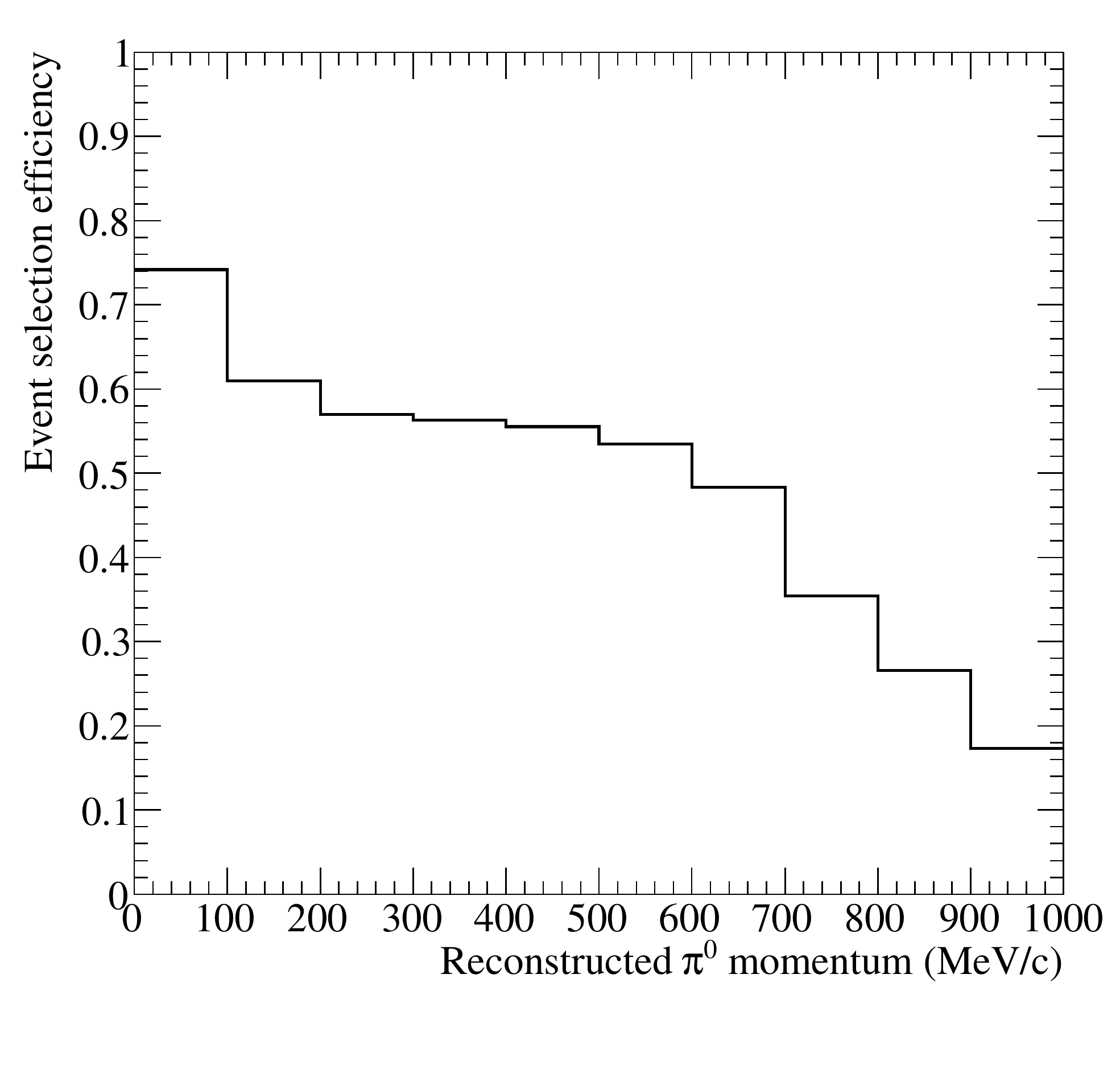} 
   \label{fig:sk1efficpi0}
 } 
 \subfigure[]{ 
    \includegraphics[width=0.4\textwidth,keepaspectratio=true,type=pdf,ext=.pdf,read=.pdf]{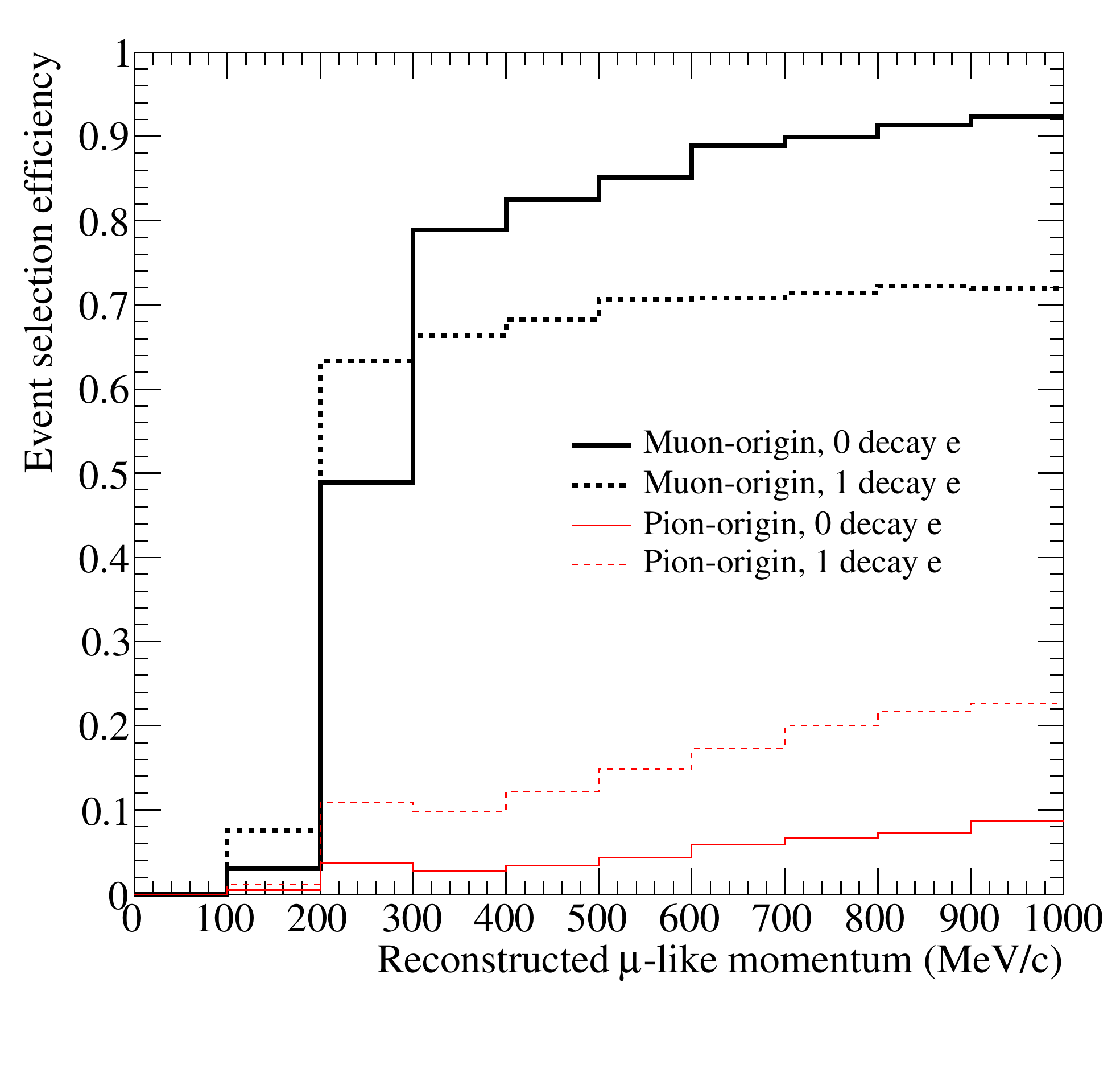} 
    \label{fig:sk1efficpip}
 } 
 \vspace{-2mm}
 \caption{ Event selection efficiencies for atmospheric neutrino MC as a function of 
           reconstructed momentum for (a)~\nnupiz selection cuts and (b)~\pnupip selection 
           cuts. For the \pip sample, all momenta are calculated assuming a muon hypothesis. 
           The atmospheric neutrino events that make up the background for the \pnupip
           mode are treated separately in the spectrum fit according to their origin. }
\end{figure*}

The event selection cuts applied to the fully-contained atmospheric neutrino
data are: (A) the number of Cherenkov rings is two for \nnupiz (one
for \pnupip), (B) all rings are showering (electron-like) for \nnupiz
(non-showering (muon-like) for \pnupip), (C) the number of electrons from
muon decay is zero for \nnupiz (zero or one for \pnupip), and (D) the
reconstructed momentum is less than 1000~\mevc. For the \nnupiz sample only,
there is one additional requirement (E) that the reconstructed invariant mass
of the \piz is between 85~\mevcc and 185~\mevcc. A discussion of the momentum
reconstruction and performance can be found in
Refs.~\cite{Nishino:2011LPM,Regis:2012muK0}. The selected fraction of single
pions in the atmospheric neutrino MC with these cuts are shown as a function
of reconstructed momentum for \ski in Fig.~\ref{fig:sk1efficpi0} for \piz,
and \ref{fig:sk1efficpip} for \pip. The efficiency curves for the \skii
and \skiii periods, not shown here but treated individually in the fit, are
very similar to the \ski curves shown in the figure. For the \piz mode, the
majority of atmospheric neutrino-induced background events that pass the
selection cuts arise from neutral current (NC) single pion production (76\%).
The other backgrounds are due to charged current (CC) single pion production
with the outgoing charged lepton below Cherenkov threshold (7\%), NC multiple
pion production (7\%), and small fractions of other processes (less than a
few percent each). As this background is overwhelmingly made up of pions and
their selection efficiency as a function of momentum is not flat, the
uncertainty in selection efficiency is treated as a systematic error binned
in momentum. Efficiency is defined as the fraction of all fully-contained
events in the fiducial volume (FCFV) with at least one \piz and no decay
electrons, which pass the selection cuts for \nnupiz. For the \pip sample, a
large fraction of the background events are of non-pionic origin, as shown in
the first row of Table~\ref{tab:bgcontribpip}.  Since the shape and level of
the selection efficiency differs for backgrounds of pionic and muonic origin,
as demonstrated in Fig.~\ref{fig:sk1efficpip}, these two cases are treated as
separate systematic errors in the spectrum fit. Efficiency curves in this
case are calculated using the MC truth neutrino interaction mode in the
denominator to determine what fraction of events pass the \pnupip selection
cuts for: (1) FCFV events which originate from charged current quasi-elastic
interactions (muonic origin), and (2) FCFV events which originate from
interactions that truly have a \pip (pionic origin).

\begin{table}[htb]
\begin{center}
\caption{Sources of atmospheric neutrino- and antineutrino-induced background 
for the \pnupip decay mode.}  
  \begin{tabular}{lc}
  \hline
  \hline
  \vspace{0.5mm}
  Interaction mode   &  Fraction of background \\
  \hline
   CCQE ($\nu n\rightarrow \mu p$)                        &   0.75 \\
   Single-$\pi$ ($\nu N \rightarrow \ell N' \pi^{(+/-/0)}$)   &   0.20 \\
   Multi-$\pi$ ($\nu N \rightarrow \ell N' (n\pi)$)           &   0.03 \\
   Other                                                      &   0.02 \\
  \hline 
  \hline 
  \end{tabular} 
   \label{tab:bgcontribpip}
\end{center}
\end{table}

The SK particle identification (PID) algorithm only classifies Cherenkov
rings as showering ($e^{\pm},\gamma$) or non-showering
($\mu^{\pm},\pi^{\pm}$), as described in Ref.~\cite{Ashie:2005}. Another PID
algorithm which uses additional information about the Cherenkov ring opening
angle to attempt further classification between $\mu$ and $\pi^{\pm}$ rings
also exists~\cite{Nishino:2011LPM}, but was shown to add no improvement to
this analysis; thus, no attempt is made to separate muon-induced rings from
pion-induced rings.


The fit to the reconstructed momentum spectrum is a $\chi^{2}$ minimization
based on a Poisson probability with systematic errors taken into account by
quadratic penalties (pull terms). This technique is the same as that
described by Fogli {\em{et al.}}, in Ref.~\cite{Fogli:2002}. The $\chi^{2}$ is
defined as

%
\begin{equation}
\begin{split}
\chi^{2} & = 2 \sum^{{\textrm{nbins}}}_{i=1} \left(  N^{{\textrm{exp}}}_{i}( 1 + \sum^{N_{{\textrm{syserr}}}}_{j=1} f^{j}_{i} \epsilon_{j} ) - N^{{\textrm{obs}}}_{i} \right.\\
            &  \left. +~N^{{\textrm{obs}}}_{i} \ln \frac{ N^{{\textrm{obs}}}_{i} }{ N^{{\textrm{exp}}}_{i}( 1 + \sum^{N_{{\textrm{syserr}}}}_{j=1} f^{j}_{i} \epsilon_{j} ) } \right)\\ 
            &  + \sum^{N_{{\textrm{syserr}}}}_{j=1} \left( \frac{ \epsilon_{j} }{ \sigma_{j} } \right)^{2}, 
\label{eq:fullchi}
\end{split}
\end{equation}

\noindent 
where $i$ indexes the data bins, $N_{i}^{{\textrm{exp}}}$ is the MC
expectation, and $N_{i}^{{\textrm{obs}}}$ is the number of observed events in
the $i^{th}$ bin. The Monte Carlo simulation expectation is given by
$N_{i}^{{\textrm{exp}}}= \alpha\cdot N_{i}^{{\textrm{bkg}}}+\beta\cdot
N_{i}^{{\textrm{sig}}}$, where $\alpha$ and $\beta$ are the normalization
parameters for background (atmospheric neutrinos) and signal (nucleon decay),
respectively. In the two-dimensional fit space, the parameter allowed ranges
are defined such that a value of 1.0 for $\alpha$ (atmospheric neutrino
background normalization) and a value of 0.0 for $\beta$ (nucleon decay
signal event normalization) would indicate that the SK data are perfectly
described by the atmospheric neutrino simulation alone, with no contribution
from nucleon decay. The effect of the $j^{th}$ systematic error is included
via a ``pull term'' which includes the error parameter $\epsilon_{j}$ and
$f^{j}_{i}$, which is the fractional change in the MC expectation for bin $i$
that would occur for a 1-sigma change in systematic error $\sigma_{j}$. In
total, 30 bins are used to compute the value of $\chi^{2}$ for the \nnupiz
analysis (10 for \ski, 10 for \skii, and 10 for \skiii), and 60 bins are used
for the \pnupip analysis. The number of bins in the \pip analysis is double
that of the \piz analysis due to the presence of two independent subsamples
(events with zero and one decay electron, respectively) in the selected data
for that decay mode.

Equation~\ref{eq:fullchi} is minimized with respect to the $\epsilon_{j}$
according to $\partial\chi^{2}/\partial\epsilon_{j}=0$, which yields a set of
iteratively solved equations in the epsilons. The $\chi^{2}$ is calculated by
this procedure for 10,000 points in the fit parameter space
($\alpha:[0.8,1.2], \beta:[0.0,0.2])$.  Each SK run period uses an
independent sample of 500 years of atmospheric neutrino MC simulation, and an
independent sample of 5000 events of the pertinent nucleon decay MC
simulation to calculate $N_{i}^{{\textrm{exp}}}$. The global minimum
$\chi^{2}$ for each decay mode is defined as that decay mode's best fit
point.

\begin{figure}[htb]
\includegraphics[width=0.4\textwidth,keepaspectratio=true,type=pdf,ext=.pdf,read=.pdf]{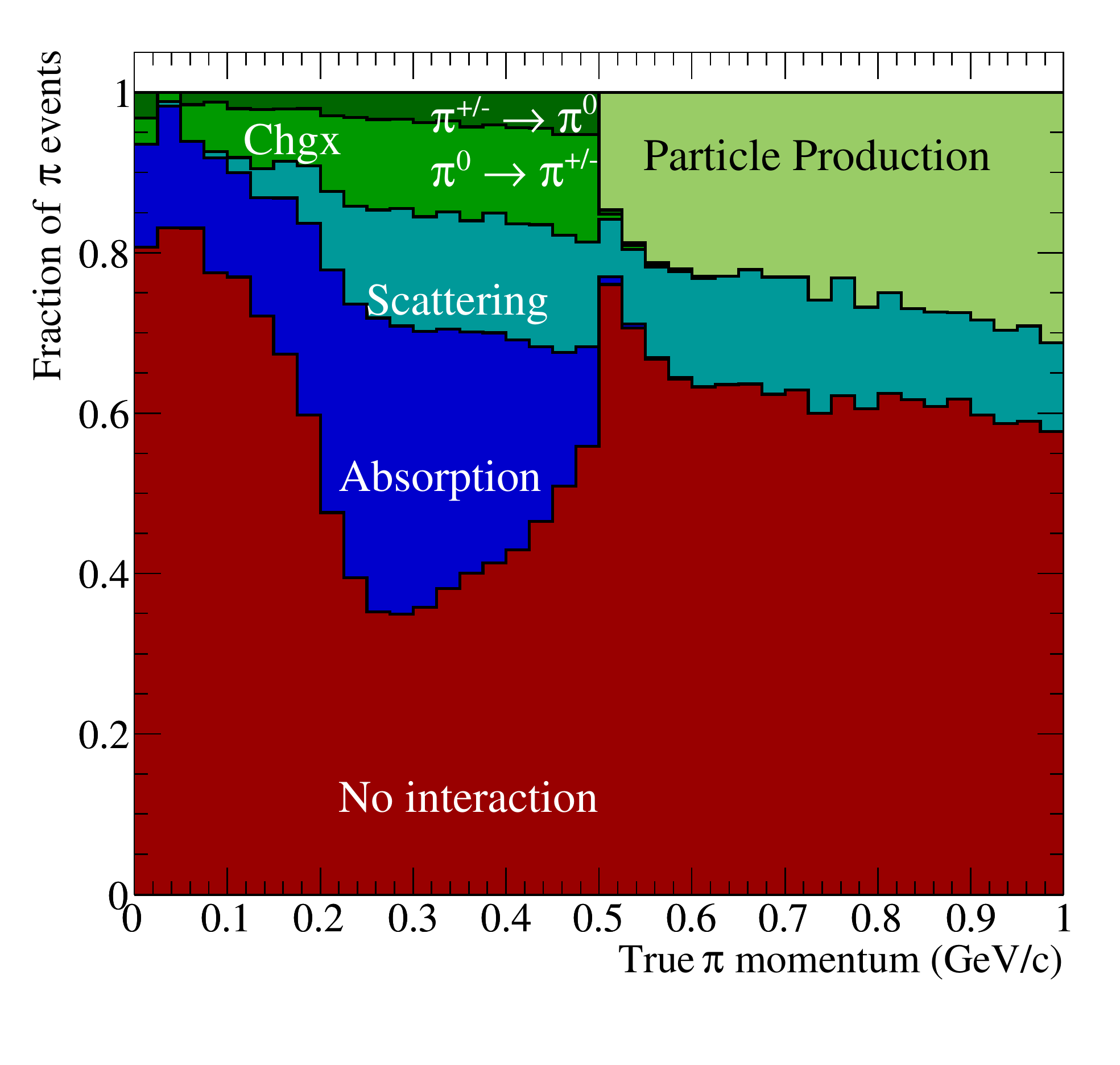}
\vspace{-5mm}
\caption{ Cumulative fractions of nuclear effects for \piz as a function of true 
          \piz momentum for atmospheric neutrino interactions. The fractions
          of events which undergo charge exchange, multiple particle
          production, scattering, and absorption are shown by various shades
          as labeled. Pions which exit the nucleus without experiencing any
          nuclear effect are indicated by the portion labeled ``No
          interaction.''  }
\label{fig:nuceff}
\end{figure}

Six sources of systematic uncertainty are considered in the \nnupiz analysis,
and 15 sources are considered in the \pnupip analysis. These can be divided
into two classes: those that are common to all SK run periods, and those that
depend on the detector geometry of an individual run period. Uncertainties in
the nuclear effect cross sections (charge exchange and particle production,
absorption, and inelastic scattering) are dominant and are treated as common
to all SK run periods. We neglect other common uncertainties such as
atmospheric neutrino flux and neutrino interaction cross sections because
they are overwhelmed by the nuclear effect uncertainties. Uncertainties in
detection efficiency are treated independently for each subsample and run
period. The detection efficiencies for \nnupiz nucleon decay signal, for \piz
background, and for muon-induced background in the \pnupip analysis, are
taken to have an overall $5\%$ uncertainty due to contributions from detector
performance uncertainties, as described in Ref.~\cite{Nishino:2011LPM}. The
signal and background detection efficiency uncertainties in each decay mode
are treated as fully correlated. For \pnupip signal and \pip-induced
backgrounds, the detection efficiency uncertainty is estimated to be larger
($15\%$) due to the possibility that charged pions may interact hadronically
as they travel through the water.

The systematic errors that contribute the most to the fits are those from
nuclear effects.  The cumulative fractions of each category of nuclear effect
are shown in Fig.~\ref{fig:nuceff} as a function of true \piz momentum for
atmospheric neutrino events (and the corresponding plot for \pip momentum is
similar). We assume an uncertainty of $30\%$ on the cross section for each
nuclear effect. As is seen in the figure, charge exchange and absorption
effects occur with greater frequency in the momentum range of these decay
modes ($\sim460$~\mevc). The discontinuity at 500~\mevc is the result of the
transition from the custom simulation used to track pions in the nucleus that
are below 500~\mevc and the GCALOR simulation of pion propagation for pions
with momentum equal to or above 500~\mevc. We anticipate that future SK
analyses will model this transition region more completely, but it is well
covered by the systematic errors in the current analysis.

The systematic errors used in this analysis and their uncertainties and
relative pulls after performing the fitting procedure are shown in
Table~\ref{tab:syserr}. As can be seen in the table, all of the systematic
error pulls stay near or below 1$\sigma$ of their nominal values afer the
fit, indicating no strong tension between data and MC simulation.

\begin{table*}[htb]
\begin{center}
\caption{Systematic error terms in the \nnupiz and \pnupip spectrum fits, with $1\sigma$
         uncertainties and resulting size of pull terms after fit.}  
  \begin{tabular}{llcc} \hline \hline
   Decay mode & Systematic error   & $1\sigma$ uncertainty & Size of pull after fit\\
        &      &         (\%)          &  (units of $\sigma$) \\
  \hline
  \multirow{6}{*}{\nnupiz}
   & Charge exchange + Particle production cross section & 30   &  -1.26 \\
   & Pion absorption cross section                       & 30   &   0.83 \\
   & Inelastic scattering cross section                  & 30   &  -0.42 \\
   & \ski single \piz BG detection efficiency            & 5    &  0.08 \\
   & \skii  single \piz BG detection efficiency          & 5    &  -0.19 \\
   & \skiii single \piz BG detection efficiency          & 5    &  -0.01 \\
  \hline
  \multirow{9}{*}{\pnupip} 
   & Charge exchange + Particle production cross section & 30      &   $-0.61$ \\
   & Pion absorption cross section                       & 30      &   $0.30$ \\
   & Inelastic scattering cross section                  & 30      &   $-0.12$ \\
   & \ski muonic BG detection efficiency 0 decay-e (1 decay-e)     &   $5~(5)$   &  $0.01~(0.16)$ \\
   & \skii muonic BG detection efficiency 0 decay-e (1 decay-e)      & $5~(5)$   &  $0.00~(0.25)$ \\
   & \skiii muonic BG detection efficiency 0 decay-e (1 decay-e)     & $5~(5)$   &  $-0.01~(0.18)$ \\
   & \ski pionic BG detection efficiency 0 decay-e (1 decay-e)       & $15~(15)$ &  $0.01~(-0.29)$ \\
   & \skii pionic BG detection efficiency 0 decay-e (1 decay-e)      & $15~(15)$ &  $0.00~(-0.08)$ \\
   & \skiii pionic BG detection efficiency 0 decay-e (1 decay-e)     & $15~(15)$ &  $0.00~(0.01)$ \\
  \hline 
  \hline 
  \end{tabular} 
   \label{tab:syserr}
\end{center}
\end{table*}

\begin{table*}[htb]
\begin{center}
\caption{Best fit parameter values, signal detection efficiency 
         for each SK running period, 90\%~C.L. value of $\beta$ parameter,
         and allowed number of nucleon decay events in the full 172.8~\ktyrs
         exposure (\ski: 91.7, \skii: 49.2, \skiii: 31.9), and lower partial
         lifetime limit for each decay mode at 90\%
           C.L.}  \begin{tabular}{lcccccc} \hline \hline
  Decay mode   &  Best fit values   & Unphysical best fit & Signal efficiency & $\beta_{{\mathrm{90CL}}}$ &  Num. signal events  &  $\tau/{\mathcal{B}}$ \\ 
               &  $(\alpha,\beta)$  & $(\alpha,\beta)$ &  (\ski, -II, -III)\%   & & at 90\%~C.L. ($N_{{\mathrm{90CL}}}$) & ($\times10^{32}$~yrs)\\
  \hline
   \nnupiz & $(0.940,0.0)$ & (0.976,-0.02) &  (48.5, 44.0, 48.5) & 0.02 & 19.1 & 11.0\\
   \pnupip & $(0.976,0.0)$ & (0.996,-0.01) &  0-decay-e: (20.4, 23.0, 22.0) & 0.01 & 52.8 & 3.9\\
           &               &               &  1-decay-e: (14.8, 12.4, 14.2) &      &      &    \\
  \hline 
  \hline 
  \end{tabular} 
   \label{tab:results}
\end{center}
\end{table*}

\begin{figure*}[h!]
   \includegraphics[width=0.4\textwidth,keepaspectratio=true,type=pdf,ext=.pdf,read=.pdf]{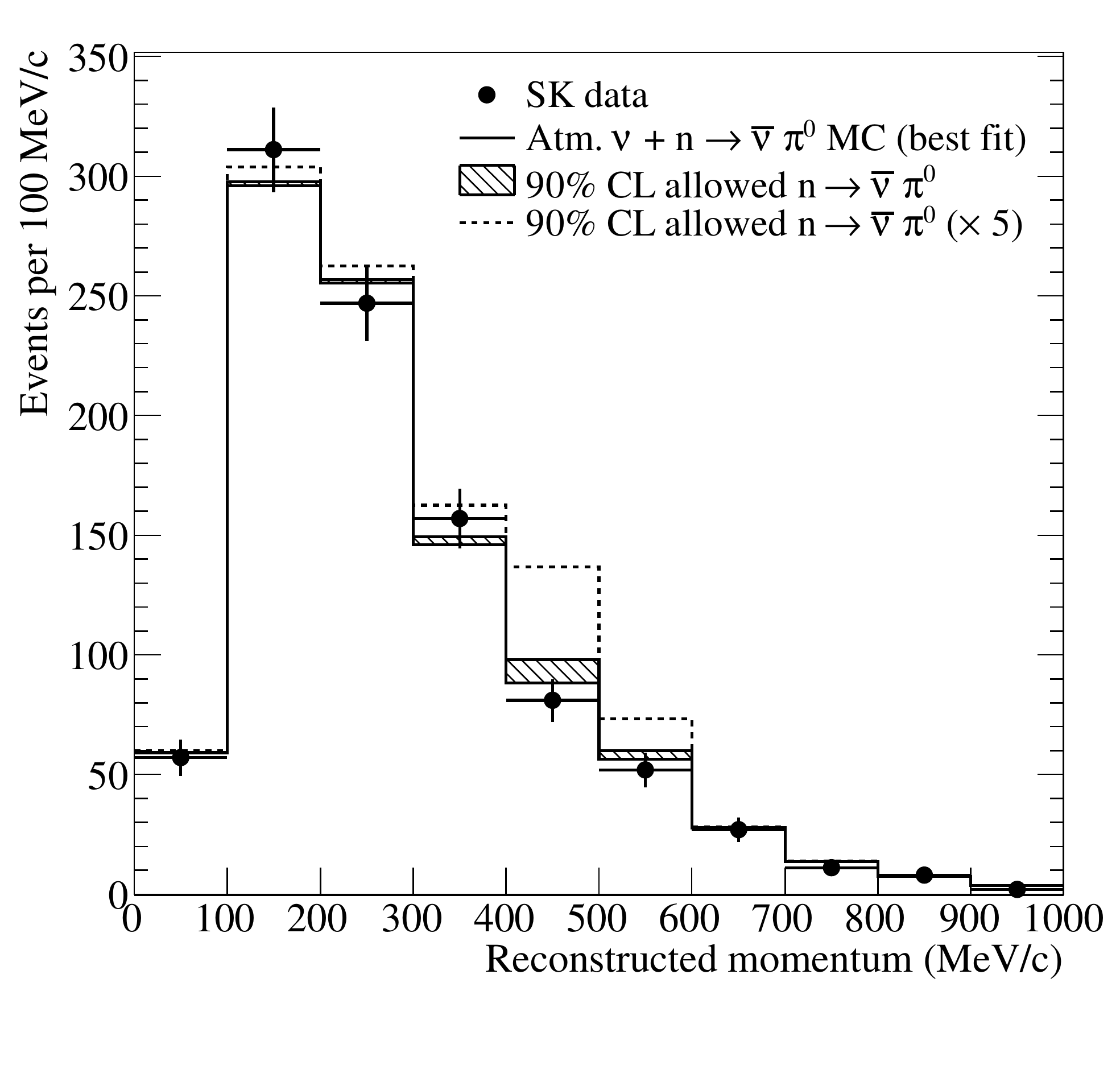} 
    \includegraphics[width=0.4\textwidth,keepaspectratio=true,type=pdf,ext=.pdf,read=.pdf]{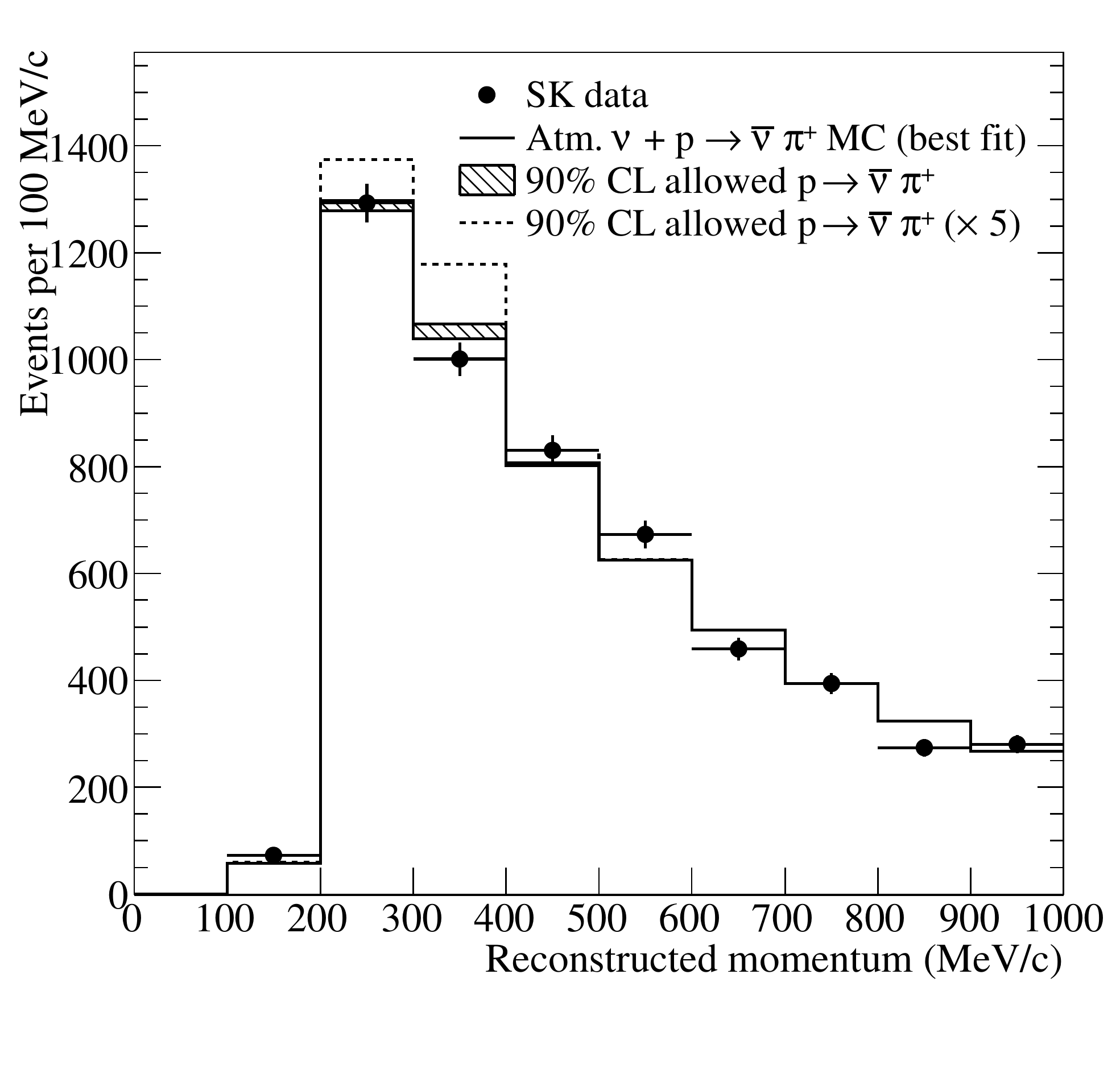} 
  \vspace{-5mm}
 \caption{ Reconstructed momentum for 172.8 \ktyrs of \ski+II+III
    data (black dots), best fit result of atmospheric neutrino plus nucleon
    decay MC simulation (solid line), and the 90\% C.L. allowed amount of
    nucleon decay (hatched histogram) for \nnupiz (left) and \pnupip (right).
    The dashed line shows how a positive signal of nucleon decay would look, 
    corresponding to 5 times the limit we set on the decay partial lifetimes. 
    The \pnupip nucleon decay contribution in the right figure is reconstructed 
    at lower momentum than the expected value (458.8~\mevc) because a muon 
    hypothesis is assumed in the reconstruction.}
\label{fig:finalfit}
\end{figure*}

In an unconstrained fit, the best fit value for the $\beta$ parameter
(nucleon decay normalization) falls in the unphysical region for both
the \nnupiz and the \pnupip analyses, preferring a negative amount of nucleon
decay. To avoid this, we constrain the $\beta$ values to be in the physical
region, and determine the $90\%$ confidence level (C.L.) value of $\beta$
according to the Feldman-Cousins prescription~\cite{Feldman:1998p3351}. The
partial lifetime lower limit for each decay mode is then calculated according
to
\begin{equation}
\tau/\mathcal{B}=\frac{\Delta t \cdot \varepsilon \cdot N_{{\textrm{nucleons}}}}{N_{{\textrm{90CL}}}},
\end{equation}
where $\mathcal{B}$ is the decay mode branching ratio, $\Delta t$ is the
exposure in \ktyrs, $\varepsilon$ is the overall signal detection efficiency of the
nucleon decay mode, $N_{{\textrm{nucleons}}}$ is the number of nucleons per kiloton of
water ($2.7\times10^{32}$ neutrons or $3.3\times10^{32}$ protons), and
$N_{{\textrm{90CL}}}$ is the number of nucleon decay events allowed at 90\%~C.L. as
determined from the $\beta$ value.

The physical and unphysical best fit parameter values, signal detection
efficiencies for each running period of SK, and number of nucleon decay
events allowed in the data sample at $90\%$~C.L. are shown in
Table~\ref{tab:results}. Using the constrained physical best fit parameters
and pull terms, the resulting fitted momentum spectra are shown combined for
all running periods as the solid black lines in Fig.~\ref{fig:finalfit}, with
the non-observation 90\%~C.L. allowed amount of signal nucleon decay shown by
the hatched histogram and overlaid SK data by black dots. The
90\%~C.L. partial lifetime lower limits we set for these two decay modes are
then $\tau_{\nnupiz}>1.1\times10^{33}$~years and
$\tau_{\pnupip}>3.9\times10^{32}$~years. In comparison, the predicted range
of partial lifetimes allowed for the SO(10) model presented in
Ref.~\cite{Goh:2004p279} are
${\tau_{\nnupiz}}=2{\tau_{\pnupip}}\leq5.7-13\times10^{32}$~years; the
model's allowed ranges are nearly ruled out by the limits presented
here. These limits represent an order of magnitude improvement over
previously published limits for these two decay
modes~\cite{McGrew:1999p500,Hirata:1989p2467,Wall:2000p3359} and can be used
to more tightly constrain other GUT models that allow these modes.

\begin{acknowledgments}
We gratefully acknowledge cooperation of the Kamioka Mining and Smelting
Company. The Super-Kamiokande experiment was built and has been operated with
funding from the Japanese Ministry of Education, Science, Sports and Culture,
and the United States Department of Energy.
\end{acknowledgments}

\bibliographystyle{apsrev}
\bibliography{references}

\begin{thebibliography}{16}
\expandafter\ifx\csname natexlab\endcsname\relax\def\natexlab#1{#1}\fi
\expandafter\ifx\csname bibnamefont\endcsname\relax
  \def\bibnamefont#1{#1}\fi
\expandafter\ifx\csname bibfnamefont\endcsname\relax
  \def\bibfnamefont#1{#1}\fi
\expandafter\ifx\csname citenamefont\endcsname\relax
  \def\citenamefont#1{#1}\fi
\expandafter\ifx\csname url\endcsname\relax
  \def\url#1{\texttt{#1}}\fi
\expandafter\ifx\csname urlprefix\endcsname\relax\def\urlprefix{URL }\fi
\providecommand{\bibinfo}[2]{#2}
\providecommand{\eprint}[2][]{\url{#2}}

\bibitem[{\citenamefont{Georgi and Glashow}(1974)}]{Glashow:1974p1461}
\bibinfo{author}{\bibfnamefont{H.}~\bibnamefont{Georgi}} \bibnamefont{and}
  \bibinfo{author}{\bibfnamefont{S.~L.} \bibnamefont{Glashow}},
  \bibinfo{journal}{Phys Rev Lett} \textbf{\bibinfo{volume}{32}},
  \bibinfo{pages}{438} (\bibinfo{year}{1974}).

\bibitem[{\citenamefont{Pati and Salam}(1974)}]{Salam:1974p1424}
\bibinfo{author}{\bibfnamefont{J.~C.} \bibnamefont{Pati}} \bibnamefont{and}
  \bibinfo{author}{\bibfnamefont{A.}~\bibnamefont{Salam}},
  \bibinfo{journal}{Phys Rev D} \textbf{\bibinfo{volume}{10}},
  \bibinfo{pages}{275} (\bibinfo{year}{1974}).

\bibitem[{\citenamefont{Goh et~al.}(2004)\citenamefont{Goh, Mohapatra, Nasri,
  and Ng}}]{Goh:2004p279}
\bibinfo{author}{\bibfnamefont{H.}~\bibnamefont{Goh}},
  \bibinfo{author}{\bibfnamefont{R.}~\bibnamefont{Mohapatra}},
  \bibinfo{author}{\bibfnamefont{S.}~\bibnamefont{Nasri}}, \bibnamefont{and}
  \bibinfo{author}{\bibfnamefont{S.}~\bibnamefont{Ng}}, \bibinfo{journal}{Phys
  Lett B} \textbf{\bibinfo{volume}{587}}, \bibinfo{pages}{105}
  (\bibinfo{year}{2004}).

\bibitem[{\citenamefont{Babu et~al.}(2010)\citenamefont{Babu, Pati, and
  Tavartkiladze}}]{Babu:2010p1621}
\bibinfo{author}{\bibfnamefont{K.}~\bibnamefont{Babu}},
  \bibinfo{author}{\bibfnamefont{J.}~\bibnamefont{Pati}}, \bibnamefont{and}
  \bibinfo{author}{\bibfnamefont{Z.}~\bibnamefont{Tavartkiladze}},
  \bibinfo{journal}{JHEP} \textbf{\bibinfo{volume}{06}}, \bibinfo{pages}{1}
  (\bibinfo{year}{2010}).

\bibitem[{\citenamefont{Fukuda et~al.}(2003)}]{Fukuda:2003p1333}
\bibinfo{author}{\bibfnamefont{S.}~\bibnamefont{Fukuda}} \bibnamefont{et~al.},
  \bibinfo{journal}{Nucl Inst Meth} \textbf{\bibinfo{volume}{A}}
  (\bibinfo{year}{2003}).

\bibitem[{\citenamefont{Nishino et~al.}(2012)}]{Nishino:2011LPM}
\bibinfo{author}{\bibfnamefont{H.}~\bibnamefont{Nishino}} \bibnamefont{et~al.},
  \bibinfo{journal}{Phys Rev D}  (\bibinfo{year}{2012}),
  \eprint{hep-ex/1203.4030}.

\bibitem[{\citenamefont{Hayato}(2002)}]{neut}
\bibinfo{author}{\bibfnamefont{Y.}~\bibnamefont{Hayato}},
  \bibinfo{journal}{Nucl Phys B Proc Suppl} \textbf{\bibinfo{volume}{112}},
  \bibinfo{pages}{171} (\bibinfo{year}{2002}).

\bibitem[{\citenamefont{Honda et~al.}(2007)}]{Honda:2007}
\bibinfo{author}{\bibfnamefont{M.}~\bibnamefont{Honda}} \bibnamefont{et~al.},
  \bibinfo{journal}{Phys Rev D} \textbf{\bibinfo{volume}{75}},
  \bibinfo{pages}{043006} (\bibinfo{year}{2007}).

\bibitem[{Gea(1993)}]{Geant3}
\bibinfo{journal}{CERN Program Library Long Writeup W5013}
  (\bibinfo{year}{1993}).

\bibitem[{\citenamefont{Ashie et~al.}(2005)}]{Ashie:2005}
\bibinfo{author}{\bibfnamefont{Y.}~\bibnamefont{Ashie}} \bibnamefont{et~al.},
  \bibinfo{journal}{Phys Rev D} \textbf{\bibinfo{volume}{71}},
  \bibinfo{pages}{112005} (\bibinfo{year}{2005}).

\bibitem[{\citenamefont{Regis et~al.}(2012)}]{Regis:2012muK0}
\bibinfo{author}{\bibfnamefont{C.}~\bibnamefont{Regis}} \bibnamefont{et~al.},
  \bibinfo{journal}{Phys Rev D} \textbf{\bibinfo{volume}{86}}
  (\bibinfo{year}{2012}), \eprint{hep-ex/1205.6538}.

\bibitem[{\citenamefont{Fogli et~al.}(2002)\citenamefont{Fogli, Lisi
  et~al.}}]{Fogli:2002}
\bibinfo{author}{\bibfnamefont{G.~L.} \bibnamefont{Fogli}},
  \bibinfo{author}{\bibfnamefont{E.}~\bibnamefont{Lisi}}, \bibnamefont{et~al.},
  \bibinfo{journal}{Phys Rev D} \textbf{\bibinfo{volume}{66}},
  \bibinfo{pages}{053010} (\bibinfo{year}{2002}).

\bibitem[{\citenamefont{Feldman and Cousins}(1998)}]{Feldman:1998p3351}
\bibinfo{author}{\bibfnamefont{G.~J.} \bibnamefont{Feldman}} \bibnamefont{and}
  \bibinfo{author}{\bibfnamefont{R.~D.} \bibnamefont{Cousins}},
  \bibinfo{journal}{Phys Rev D} \textbf{\bibinfo{volume}{57}},
  \bibinfo{pages}{3873} (\bibinfo{year}{1998}).

\bibitem[{\citenamefont{McGrew et~al.}(1999)}]{McGrew:1999p500}
\bibinfo{author}{\bibfnamefont{C.}~\bibnamefont{McGrew}} \bibnamefont{et~al.},
  \bibinfo{journal}{Phys Rev D} \textbf{\bibinfo{volume}{59}},
  \bibinfo{pages}{052004} (\bibinfo{year}{1999}).

\bibitem[{\citenamefont{Hirata et~al.}(1989)}]{Hirata:1989p2467}
\bibinfo{author}{\bibfnamefont{K.~S.} \bibnamefont{Hirata}}
  \bibnamefont{et~al.}, \bibinfo{journal}{Phys Lett B}
  \textbf{\bibinfo{volume}{220}}, \bibinfo{pages}{308} (\bibinfo{year}{1989}).

\bibitem[{\citenamefont{Wall et~al.}(2000)}]{Wall:2000p3359}
\bibinfo{author}{\bibfnamefont{D.}~\bibnamefont{Wall}} \bibnamefont{et~al.},
  \bibinfo{journal}{Phys Rev D} \textbf{\bibinfo{volume}{62}},
  \bibinfo{pages}{092003} (\bibinfo{year}{2000}).

\end{thebibliography}

\end{document}